\documentclass[runningheads]{llncs}

\usepackage[T1]{fontenc}
\usepackage{amsfonts}
\usepackage{graphicx}
\usepackage{hyperref}
\usepackage{color}
\usepackage[numbers,square,sort&compress]{natbib}

\usepackage{acronym}
\usepackage{xspace}
\usepackage{amsmath}
\usepackage{booktabs}
\usepackage[inline]{enumitem}
\usepackage{footmisc}
\usepackage{multirow}
\usepackage{caption}
\usepackage{subcaption}

\newcommand{\sigmoid}[1]{\operatorname{sigmoid}(#1)}
\newcommand{\ReLU}[1]{\operatorname{ReLU}(#1)}

\newcommand{\bo}{\mathbf{o}}
\newcommand{\bh}{\mathbf{h}}
\newcommand{\bp}{\mathbf{p}}
\newcommand{\bm}{\mathbf{m}}
\newcommand{\bq}{\mathbf{q}}
\newcommand{\bW}{\mathbf{W}}
\newcommand{\boldb}{\mathbf{b}}
\newcommand{\itemset}[1]{\mathcal{I}_{#1}}
\newcommand{\userset}[1]{\mathcal{U}_{#1}}
\newcommand{\mkt}[1]{\mathbb{M}_{#1}}
\newcommand{\useru}[2]{U_{#1}^{#2}}

\newcommand{\partitle}[1]{\vspace{0.05in} \noindent \textbf{#1}}

\newcommand\topic[1]{{#1}}

\newcommand\resulthl[1]{{\textit{#1}}}

\newcommand{\avgsrc}{\texttt{AVG}~}
\newcommand{\bestsrc}{\texttt{BST}~}

\acrodef{MTL}{multi-task learning}
\acrodef{CF}{collaborative filtering}
\acrodef{CDR}{cross-domain recommendation}
\acrodef{CMR}{cross-market recommendation}
\acrodef{MAML}{model-agnostic meta-learning}
\acrodef{HR}{hit-rate}
\acrodef{MA}{market-aware}

\acrodef{GMF}{generalized matrix factorization}
\acrodef{MLP}{multi-layer perceptron}
\acrodef{NMF}{neural matrix factorization}

\acrodef{MA-GMF}{market-aware generalized matrix factorization}
\acrodef{MA-MLP}{market-aware multi-layer perceptron}
\acrodef{MA-NMF}{market-aware neural matrix factorization}

\acrodef{MF}{matrix factorization}

\begin{document}
\title{Market-Aware Models for Efficient Cross-Market Recommendation}

\author{Samarth Bhargav\orcidID{0000-0001-5204-8514} \and
Mohammad Aliannejadi\orcidID{0000-0002-9447-4172} \and
Evangelos Kanoulas\orcidID{0000-0002-8312-0694}}

\institute{University of Amsterdam, Amsterdam, The Netherlands 
\email{\{s.bhargav,m.aliannejadi,e.kanoulas\}@uva.nl}}

\authorrunning{S. Bhargav et al.}

\maketitle   

\begin{abstract}
We consider the \acf{CMR} task, which involves recommendation in a low-resource \textit{target} market using data from a richer, auxiliary \textit{source} market. Prior work in \ac{CMR} utilised meta-learning to improve recommendation performance in target markets; meta-learning however can be complex and resource intensive. In this paper, we propose \ac{MA} models, which directly model a market via \textit{market embeddings} instead of meta-learning across markets. These embeddings transform item representations into market-specific representations. Our experiments highlight the effectiveness and efficiency of \ac{MA} models both in a pairwise setting with a single target-source market, as well as a global model trained on all markets in unison. In the former pairwise setting, \ac{MA} models on average outperform market-unaware models in 85\% of cases on nDCG@10, while being time-efficient --- compared to meta-learning models, \ac{MA} models require only ~15\% of the training time. In the global setting, \ac{MA} models outperform market-unaware models consistently for some markets, while outperforming meta-learning-based methods for all but one market. We conclude that \ac{MA} models are an efficient and effective alternative to meta-learning, especially in the global setting.

\end{abstract}

\section{Introduction}

\topic{\Acf{CMR} involves improving recommendation performance in a target market using data from one or multiple auxiliary source markets}. Data from \textit{source} markets, 
which have rich- and numerous interactions, are leveraged to aid performance in a \textit{target} market with fewer interactions. For instance, an e-commerce company well-established in Germany may want to start selling its products in Denmark. Using \ac{CMR} methods, data from the German market can be utilised to augment recommender performance in the Danish market. This task is challenging since target market data can be scarce or otherwise unavailable, and user behaviours may differ across markets \cite{xmrec_paper, roitero_leveraring2020, ferwerda2016exploring}. %

Research in \ac{CMR} tackles multiple challenges. One challenge is to select the best source market, which is crucial since user behaviours across markets may vary \cite{roitero_leveraring2020, xmrec_paper}, which may harm performance instead of bolstering it. Furthermore, effectively utilising data from multiple markets at the same time without harming performance can be challenging \cite{xmrec_paper}. Another key obstacle is effectively modelling a market, in addition to users and items. \citet{xmrec_paper} treat recommendation in each market as a task in a \ac{MTL} framework, using meta-learning to learn model parameters. This is followed by a fine-tuning step per market. These two steps enable models to learn both common behaviours across markets as well as market-specific behaviours. However, meta-learning can be resource intensive compared to other methods. In addition to this, utilising new data from source markets requires re-running the meta-learning step.

We propose \acf{MA} models to address these limitations. We aim to \textit{explicitly} model each market as an embedding, using which an item representation can be transformed and `customised' for the given market. Compared to meta-learning models, we show that \ac{MA} models are far more efficient to train. Furthermore, they are trained in one go, enabling easier model updates when new data is collected. \ac{MA} models are built on the hypothesis that explicit modelling of markets allows better generalisation. In essence, an item representation is a product of 
\begin{enumerate*}[label=(\roman*)]
    \item an \textit{across-market} item embedding and 
    \item a market embedding.
\end{enumerate*} 
The former is learnt from data across markets, and aims to capture an item representation applicable across markets; the latter enables market-specific behaviours to be captured.

In our experiments, we compare \ac{MA} models with market-unaware baselines as well as meta-learning models. We do so in multiple settings, utilising data from several markets: the \textit{pairwise} setting, which deals with a single target-source pair, and the \textit{global} setting which trains one model for recommendation in all markets.  In the \textit{pairwise} setting, we show that \textit{MA} models improve over market-unaware models for many markets, and match or beat meta-learning methods. This is significant since we show that training \ac{MA} models require approximately the same time as market-unaware models and only 15\% of the time required to train meta-learning models. We show that \ac{MA} models especially excel in the \textit{global} setting, outperforming meta-learning methods for nearly every market. We examine the following research questions\footnote{\url{https://github.com/samarthbhargav/efficient-xmrec}\label{opensource}}:

\begin{itemize}[leftmargin=*, label={}]
    \item \textbf{RQ1.} \textit{Given a single source and target market, does explicitly modelling markets with embeddings lead to effective performance in the target market?} We compare \ac{MA} models against market-unaware as well as meta-learning models. We show \ac{MA} models achieve the best performance for most markets, and when a single, best source is available they match or outperform baselines for all markets.        
    \item \textbf{RQ2.} \textit{How computationally expensive are \ac{MA} models compared to market-unaware and meta-learning models?} We show that \ac{MA} models require similar training times as market-unaware models, and require fewer computational resources to train compared to meta-learning models while achieving similar or better performance.
    \item \textbf{RQ3.} \textit{How do \ac{MA} models compare against market-unaware models and meta-learning models when a \emph{global} model is trained on all markets in unison?} We show that \ac{MA} models outperform or match market-unaware baselines, outperforming meta-learning models for all but one market. 
\end{itemize}

\section{Related Work}

While both \ac{CDR} and \ac{CMR} focus on improving recommender effectiveness using data from other domains (i.e.~ item categories) or markets, they present different challenges: \ac{CDR} involves recommending items in a different domain for the same set of users, with the general assumption that the model learns from interactions of overlapping users. In \ac{CMR}, \textit{items} are instead shared across different markets, with each market having a different set of users. Interactions from auxiliary markets are leveraged to boost performance for users in the target market for a similar set of items.

\partitle{Cross-domain recommendation.} \ac{CDR} has been researched extensively \cite{im2007does, lu2013selective, elkahky_multivew2015, rafailidis_collab2017, hu2018conet, li2020ddtcdr, perera2019cngan, krishnan_ctxlinvariants20202}.  Prior approaches involve clustering-based algorithms \cite{mirbakhsh_improving2015} and weighing the influence of user preferences based on the domain \cite{rafailidis_collab2017}. \citet{lu2013selective} show that domain transfer may sometimes harm performance in the target domain. Neural approaches using similarity networks like DSSM \cite{huang_dssm2013} or transfer learning \cite{elkahky_multivew2015, hu2018conet} can be effective. DDCTR \cite{li2020ddtcdr} utilises iterative training across domains. Augmenting data with `virtual' data \cite{chae_arcf2020, perera2019cngan}, as well as considering additional sources \cite{zhao2020catn} have been shown to help. Other approaches leverage domain adaptation \cite{ganin2015unsupervised} for leveraging content for full cold-start \cite{kanagawa2019cross}, utilising adversarial approaches \cite{wang2019recsys, li2020atlrec} or formulating it as an extreme classification problem \cite{yuan2019darec}.  Our approach is inspired by contextual invariants \cite{krishnan_ctxlinvariants20202}, which are behaviours that are consistent across domains, similar to our hypothesis that there are behaviours common across markets.  

\partitle{Cross-market recommendation.} \ac{CMR} is relatively new and understudied compared to \ac{CDR}. \citet{ferwerda2016exploring} studied \ac{CMR} from the perspective of country based diversity. \citet{roitero_leveraring2020} focus on \ac{CMR} for music, investigating trade-offs between learning from local/single markets vs.~a global model, proposing multiple training strategies. \citet{xmrec_paper} release a new dataset for the Cross Market Product recommendation problem, which we utilise in our experiments. They design a meta-learning approach to transfer knowledge from a source market to a target market by freezing and forking specific layers in their models. The WSDM Cup 2022 challenge also dealt with this dataset, where most top teams utilised an ensemble of models based on different data pairs. \citet{DBLP:conf/sigir/CaoCLW22} builds on the XMRec dataset and proposes multi-market recommendation, training a model to learn intra- and inter-market item similarities. In this work, we show that meta-learning methods are expensive to train. Instead, we show that market embeddings can encode and effectively transfer market knowledge, beating or matching the performance of complex models while being much more efficient to train.

\section{Methodology}

We outline market-unaware models in Section \ref{sec:meth_baselines}, followed by market-aware models as well as meta-learning models in Section \ref{sec:meth_market_augmented}.

\partitle{Notation.} Given a set of markets $\{\mkt{0}, \mkt{1}, \dots, \mkt{t} \}$, such that market $l$ has a items $\itemset{l}$ and $z_l$ users $\userset{l} = \{\useru{l}{1} \dots \useru{l}{z_l}\}$ . We assume the \textit{base} market $\mkt{0}$ has $\itemset{0} ~\text{s.t.} ~ \itemset{0} \supset \itemset{l}$ for all $1 \leq l \leq m$. The task is to adapt a given market $\mkt{l}$ using data from other markets $\mkt{m\neq l}$ as well as data from the target market. We use $\bp_u$ for the user embedding for user $u$, $\bq_i$ for the item embedding for item $i$, and finally $\bo_l$ for the market embedding for market $l$. $y_{ui}$ and $\hat{y}_{ui}$ is the actual and predicted rating respectively. $\odot$ denotes an element-wise product.

\subsection{Market-Unaware Models}\label{sec:meth_baselines}

These models do not differentiate between users and items from different markets and are termed  \textit{market-unaware} since they do not explicitly model the market. We first outline three such models previously employed for \ac{CMR} \cite{xmrec_paper, he2017ncf}:

\begin{itemize}[leftmargin=*]
    \item \textbf{GMF}: The \ac{GMF} model computes the predicted rating $\hat{y}_{ui}$ given $\bp_u$, $\bq_i$ and parameters $\bh$:
    
    \begin{align*}
        \hat{y}_{ui} = \sigmoid{\bh^T(\bp_u \odot \bq_i)}
    \end{align*}
    
    \item \textbf{MLP}: An \ac{MLP} uses a $L$ layer fully-connected network, such that:
    \begin{align*}
        \bm_0 &= \begin{bmatrix} \bp_u \\ \bq_i \end{bmatrix} \\
        \bm_{L-1} &= \ReLU{\bW_{L-1}^T \ReLU{\dots \ReLU{\bW_1^T \bm_{0} + \boldb_1}} + \boldb_{L-1}} \\
        \hat{y}_{ui} &= \sigmoid{\bh^T \bm_{L-1}}
    \end{align*}

    \item \textbf{NMF}: \ac{NMF} combines both \ac{MLP} and \ac{GMF}. Given $\bp_u^1$, $\bq_i^1$ for the \ac{MLP}, and $\bp_i^2$, $\bq_u^2$ for \ac{GMF}, the \ac{NMF} model computes the score as follows:
    
    \begin{align*}
        \bm_{0} &= \begin{bmatrix} \bp_u^1 \\ \bq_i^1 \end{bmatrix} \\
        \bm_{MLP} &= \ReLU{\bW^T_L \ReLU{ \dots \ReLU{\bW^T_1 \bm_0 + \boldb_1}}) + \boldb_L} \\
        \bm_{GMF} &= \bp_u^2 \odot \bq_i^2 \\
        \hat{y}_{ui} &= \sigmoid{\bh^T \begin{bmatrix} \bm_{GMF} \\ \bm_{MLP} \end{bmatrix}}
    \end{align*}
    
\end{itemize}

For adapting to \ac{CMR}, different sets of users from different markets are treated similarly, and training is performed on a combined item pool resulting in a single model. During inference for a user, however, only items from that market are ranked.

\subsection{Market-Aware Models}\label{sec:meth_market_augmented}

We first discuss models proposed by \citet{xmrec_paper}, followed by our proposed methods.  

\partitle{Meta-learning baselines.} \citet{xmrec_paper} propose using meta-learning in an \ac{MTL} setting where each market is treated as a `task'. \ac{MAML} \cite{finn2017model} is employed to train the base \ac{NMF} model across markets. \ac{MAML} employs two loops for training, an inner loop that optimises a particular market, and an outer loop that optimises across markets. This makes training expensive, as we will show in our experiments. Once a MAML model is trained, the FOREC model is obtained as follows for a given source/target market: (a) the MAML model weights are copied over to a new model, `forking' it, (b) parts of the weights of the model are frozen and finally (c) the frozen model is fine-tuned on the given market. 

Both MAML and FOREC are market aware but do not \textit{explicitly} model the market i.e.~ a single item embedding is learned in MAML models for all markets, and while market adaptation is achieved through fine-tuning for FOREC, it requires maintaining separate sets of parameters, unlike the proposed \ac{MA} models.

\partitle{Market Aware Models.} Markets here are explicitly modelled by learning embeddings for each of them, in addition to user and item embeddings. A market embedding \textit{adapts} an item to the current market, which we argue is crucial for items that may be perceived differently in different markets. This aspect should be reflected in the latent representation of the item, motivating our approach. Both meta-learning and \ac{MA} models learn item representations across markets, but \ac{MA} models this explicitly via an element-wise product between a representation for an item and a market embedding. This produces item embeddings adapted to a given market. We augment the market-unaware baselines with market embeddings, producing \ac{MA} models. We leave more complex methods, for instance --- a neural network that models item/market  interactions instead of an element-wise produce --- for future work.

To obtain a \textit{market-adapted} item embedding, we first (one-hot) encode a market $l$, to obtain a market embedding $\bo_l$; the dimensionality of $\bo_l$ is the same as $\bp_u$ and $\bq_i$. The scores are computed as follows for the three proposed models: 

\begin{itemize}[leftmargin=*]
    \item \textbf{MA-GMF}: For a user $u$ in market $l$, and item $i$, we have embeddings $\bp_u$, $\bo_l$ and $\bq_i$:
    \begin{align*}
        \hat{y}_{ui} =  \sigmoid{\bh^T(\bp_{u} \odot (\bo_l \odot \bq_i))}
    \end{align*}
    
    \item \textbf{MA-MLP}: This is the same as the \ac{MLP}, with the initial embedding $\bm_0$ augmented with market information:
    $\bm_{0} = \begin{bmatrix} \bp_u \\ \bq_i \odot \bo_l \end{bmatrix}$
    
    \item \textbf{MA-NMF}: The NMF model utilises both modifications listed above. That is:
    \begin{align*}
        \bm_{GMF} &= \bp_u^2 \odot (\bo_l \odot \bq_i^2) \\
        \bm_0 &= \begin{bmatrix} \bp_u^1 \\ \bq_i^1 \odot \bo_l \end{bmatrix} 
    \end{align*}
    
\end{itemize}

These models are trained similarly to the market-unaware models, except the market is taken into consideration when making recommendations. Market embeddings are learned via backpropagation, similar to how user and item embeddings are learned, using a binary cross entropy loss \cite{he2017ncf}.

Our proposed technique adds market awareness to all the models.
Besides this, the proposed models are easier to update with new interactions compared to MAML/FOREC. While FOREC requires the expensive MAML pre-training followed by the fork and fine-tune step, \ac{MA} models simply can be trained with new interaction data. In spite of this simplicity, \ac{MA} models achieve similar performance compared to meta-learning models while requiring far lesser time to train, which we demonstrate in the following section.

\section{Experimental Setup}\label{sec:experiments}

We conduct two sets of experiments. The first set of experiments trains models with a single auxiliary source market for improving recommendation performance in a given target market. We term these \textit{pairwise} experiments since one model is trained for a given source-target market pair. The second set of experiments deals with a \textit{global} model trained on all markets in unison, with the goal of improving overall performance. We outline the dataset, evaluation, baselines, hyperparameters and training followed by a description of the experiments.

\partitle{Dataset.} We use the XMarket dataset \cite{xmrec_paper} for all experiments. XMarket is an \ac{CMR} dataset gathered from a leading e-commerce website with multiple markets.  We utilise the largest subset, `Electronics', considering the following markets (\# users, \# items, \# interactions): \textit{de} (2373/ 2210/ 22247), \textit{jp} (487/ 955 /4485), \textit{in} (239/ 470/ 2015), \textit{fr} (2396/ 1911/ 22905), \textit{ca} (5675/ 5772/ 55045), \textit{mx} (1878/ 1645/ 17095), \textit{uk} (4847/ 3302/ 44515), \textit{us} (35916/ 31125/ 364339). We consider all markets except \textit{us} as a target market, with all markets (including \textit{us}) as possible source markets. Experiments are limited to XMarket as it is the only public dataset for research in \ac{CMR}. 

\partitle{Evaluation.} The data (per market) is split into a train/validation/test set, where one left-out item from the user history is used in the validation and test set. This follows the leave-one-out strategy \cite{cheng2016wide,ge2020learning, he2017ncf, hu2018conet, kang2018self}. We extract 99 negatives per user for evaluating recommender performance in the validation/test set, following \citet{xmrec_paper}. In the pairwise experiments, the best-source market is picked based on the validation set performance. We report nDCG@10 on the test set in all results, with significance tests using a paired two-sided t-test with the Bonferroni correction. While we report only nDCG@10, we note that we observed similar trends for HR@10.

\partitle{Compared methods.} Market-aware models are denoted with an `MA-' prefix, and are compared with the following models:
\begin{itemize}[leftmargin=*]
    \item Single-market models: These are models trained only on the target market data without any auxiliary source data, see Section \ref{sec:meth_baselines}. We train all three models GMF, NMF and MLP.
    \item Cross-market models: In addition to target market data, these models are trained with either one source market (for \textit{pairwise} experiments), or all source markets (for \textit{global} experiments). Models trained with at least one source market have a `++' suffix e.g.~ GMF++ and MA-GMF++.
    \item Meta-learning models (see Section \ref{sec:meth_market_augmented}) similarly utilise data from one or more auxiliary markets: 
        \begin{itemize}
            \item MAML \cite{finn2017model, xmrec_paper}: These are models trained using MAML, with weights initialised from a trained NMF++ model \cite{xmrec_paper}.
            \item FOREC \cite{xmrec_paper}: This model uses the trained MAML model to first freeze certain parts of the network, followed by a fine-tuning step on the target market. 
        \end{itemize}
\end{itemize}

\partitle{Model hyperparameters.} We set model parameters from \cite{xmrec_paper}\footnote{\url{https://github.com/hamedrab/FOREC}}: the dimensionality of the user, item, and market embeddings are set to 8, with a 3-Layer [16, 64, 32, 16, 8] network for \ac{MLP}/\ac{NMF} models. For MAML models, we set the fast learning rate $\beta=0.1$ with 20 shots. 

\partitle{Training.} All models are trained for 25 epochs using the Adam optimiser with a batch size of 1024. We use learning rates from \cite{xmrec_paper}, for GMF we use $0.005$, for \ac{MLP} and \ac{NMF} we use $0.01$. All models also utilise an L-2 regularisation loss with $\lambda=1e-7$. The \ac{NMF} model is initialised with weights from trained \ac{GMF} and \ac{MLP} models. MAML models are trained on top of the resulting \ac{NMF} models, and FOREC models utilise the trained MAML models for the fork-and-fine-tune procedure \cite{xmrec_paper}. \ac{MA} variants use the same hyperparameters as the market-unaware models. The objective function for all models is binary cross-entropy, given positive items and 4 sampled negatives \cite{xmrec_paper, he2017ncf}. For pairwise experiments, data from the source market is (randomly) down-sampled to the target market \cite{xmrec_paper}, which ensures that models are comparable across different-sized source markets. For global models, all data is concatenated together without any down-sampling\footnote{We observed that this greatly improved performance for almost all markets.}.

\partitle{Pairwise experiments.} The first set of experiments dealing with \textbf{RQ1} and \textbf{RQ2}, which we call \textit{pairwise} (Section \ref{sec:exp_pairwise}), assumes a single auxiliary market is available for a given target market. Since there are multiple source markets, we report both the average performance in the target market \textit{across source markets} --- termed \avgsrc --- as well as performance in the target market using the \textit{best source market}, termed \bestsrc. The two tables relay different results: the average performance indicates the \textit{expected} performance of a method since the `best' source market might be unknown, or only a single source may exist; whereas the best-source results are indicative of the maximum achievable performance \textit{if} a good source market is already known (this is typically unknown \cite{roitero_leveraring2020}). 

\partitle{Global experiments.} The second set of experiments corresponding to \textbf{RQ3} utilises  data from multiple auxiliary markets at once to train a \textit{global} recommender, with the goal to achieve good performance for all markets.  We term these experiments \textit{Global} (Section \ref{sec:exp_single_model}). We describe the results of the two sets of experiments in the following section.

\section{Results and Discussion}\label{sec:results}

\subsection{Pairwise Experiments}\label{sec:exp_pairwise}

\begin{table*}[t!]
\centering
\caption{\avgsrc results: Models are first trained on a single target-source pair and performance across sources are averaged. We report the nDCG@10 on the test set, with best performance in \textbf{bold}. Significance test ($p<\frac{0.05}{9}$) results are also reported comparing MA models with market-unaware ($\ddagger$), MAML ($*$) and FOREC ($+$).}
\label{tab:pairwise-experiments-avg}
\resizebox{\columnwidth}{!}{%
\begin{tabular}{@{}llllllll@{}}
\toprule
\textbf{Method} &
  \multicolumn{1}{c}{\textbf{de}} &
  \multicolumn{1}{c}{\textbf{jp}} &
  \multicolumn{1}{c}{\textbf{in}} &
  \multicolumn{1}{c}{\textbf{fr}} &
  \multicolumn{1}{c}{\textbf{ca}} &
  \multicolumn{1}{c}{\textbf{mx}} &
  \multicolumn{1}{c}{\textbf{uk}} \\ \midrule
GMF++  & 0.2045 & 0.0916          & 0.1891 & 0.2026 & 0.1937 & 0.4204 & 0.3222          \\
MA-GMF++ &
  0.2148$^{ \ddagger}$ &
  0.1079 &
  0.2013 &
  0.2022 &
  0.2203$^{ \ddagger}$ &
  0.4283$^{ \ddagger}$ &
  0.3327$^{ \ddagger}$ \\ \midrule
MLP++  & 0.2836 & 0.1653          & 0.4376 & 0.2704 & 0.2905 & 0.5274 & 0.4346          \\
MA-MLP++ &
  0.2909$^{ \ddagger + *}$ &
  0.1741 &
  \textbf{0.4502} &
  0.2805$^{ \ddagger}$ &
  \textbf{0.3073$^{ \ddagger + *}$} &
  0.5311 &
  0.4349$^{ *}$ \\ \midrule
NMF++  & 0.2927 & \textbf{0.1826} & 0.4403 & 0.2844 & 0.2844 & 0.5367 & \textbf{0.4379} \\
MA-NMF++ &
  \textbf{0.3055$^{ \ddagger + *}$} &
  0.1824 &
  0.4471 &
  \textbf{0.2893$^{ + *}$} &
  0.3002$^{ \ddagger + *}$ &
  \textbf{0.5387$^{ + *}$} &
  0.4370$^{ *}$ \\ \midrule
MAML   & 0.2808 & 0.1770          & 0.4320 & 0.2785 & 0.2794 & 0.5288 & 0.4296          \\
FOREC  & 0.2835 & 0.1758          & 0.4345 & 0.2816 & 0.2772 & 0.5302 & 0.4330          \\ \bottomrule
\end{tabular}%
}
\vspace{-.2in}
\end{table*}

Tables~\ref{tab:pairwise-experiments-avg} and \ref{tab:pairwise-experiments-best} report the results of the \textit{pairwise} experiments, where the models only use one auxiliary market at a time. We report both \avgsrc, the average performance of models using different auxiliary markets for the same target market (Table~\ref{tab:pairwise-experiments-avg}), as well as \bestsrc, the best auxiliary market (Table~\ref{tab:pairwise-experiments-best}). The best auxiliary market is determined based on the validation set performance. Moreover, the results of the single-market baseline models are only reported in Table~\ref{tab:pairwise-experiments-best}. We first examine \textbf{RQ1}, comparing the performance of \ac{MA} models against baselines in both the \avgsrc and \bestsrc settings. We end with discussion of \textbf{RQ2}, which compares training times across models.

\subsubsection{Do \ac{MA} models improve over market unaware models on average?}\label{sec:exp_pairwise_avg} Using Table \ref{tab:pairwise-experiments-avg}, we first examine if \acs{MA} models outperform market-unaware models in the \avgsrc setting e.g.~ \acs{GMF}++ against \acs{MA-GMF}++.  We see that the \acs{MA-GMF}++  outperforms \acs{GMF}++ for every market except \textit{fr}. \acs{MA-MLP}++ outperforms \acs{MLP}++ for all markets, and \acs{MA-NMF}++ outperforms \acs{NMF}++ on all markets except \textit{jp} and \textit{uk}. For the \textit{de} and \textit{ca} markets, we see that MA models always outperform their non-MA variant. In addition, for the \textit{uk} and \textit{mx} markets, \acs{MA-GMF}++ significantly outperforms \acs{GMF}++; and for \textit{fr} we see that \acs{MA-MLP}++ significantly outperforms \acs{MLP}++. Despite large improvements in some markets e.g.~ \acs{MA-MLP}++ improves nDCG@10 by $0.12$ points over \acs{MLP}++ for \textit{in}, we do not see a significant result, which may be due to the conservative Bonferroni correction, or fewer test users for \textit{in} (requiring larger effect sizes). Overall, \ac{MA} models outperform their market unaware equivalent in 18 of 21 settings. In summary, we can conclude that \resulthl{in the \avgsrc setting, the proposed market-aware models outperform market-unaware baselines for nearly all markets}. This demonstrates the robustness of \textit{MA} models since these improvements are across multiple source markets.

\subsubsection{How do \ac{MA} models compare against meta-learning models in the \avgsrc setting?} We compare \ac{MA} models against MAML and FOREC considering \avgsrc, in Table \ref{tab:pairwise-experiments-avg}. MA-GMF++ never outperforms MAML/ FOREC, but the differences in model sizes render this comparison unfair. A fairer comparison would be with \ac{MA-NMF}++: we see that it outperforms MAML for 5 of 7 markets: \textit{de}, \textit{fr}, \textit{ca}, \textit{mx} and \textit{uk}. Additionally, FOREC is significantly outperformed by MA-NMF++ for 4 of 7 markets: \textit{de}, \textit{fr}, \textit{ca} and \textit{mx}. We note, however, that at least one MA model outperforms both MAML/FOREC for all markets, and at least one MA model \textit{significantly} outperforms MAML/FOREC for \textit{de} (both), \textit{fr} (both), \textit{ca} (both), \textit{mx} (MAML only) and \textit{uk} (MAML only). Therefore, \resulthl{we can thus conclude that market-aware models either match or outperform meta-learning models for many markets in \avgsrc setting}.

\begin{table}[t!]
\centering
\caption{\bestsrc: Models are trained on all source markets, the best source is selected based on validation set performance. We report nDCG@10 on the test set, along with significance test results ($p<\frac{0.05}{12}$) comparing MA models with single market ($\dagger$), market unaware ($\ddagger$), MAML ($*$) and FOREC ($+$).}
\label{tab:pairwise-experiments-best}
\resizebox{\columnwidth}{!}{%
\begin{tabular}{@{}llllllll@{}}
\toprule
\textbf{Method} &
  \multicolumn{1}{c}{\textbf{de}} &
  \multicolumn{1}{c}{\textbf{jp}} &
  \multicolumn{1}{c}{\textbf{in}} &
  \multicolumn{1}{c}{\textbf{fr}} &
  \multicolumn{1}{c}{\textbf{ca}} &
  \multicolumn{1}{c}{\textbf{mx}} &
  \multicolumn{1}{c}{\textbf{uk}} \\ \midrule
GMF    & 0.2574 & 0.0823          & 0.0511 & 0.2502 & 0.2566 & 0.5066 & 0.4136 \\
GMF++  & 0.2670 & 0.1093          & 0.2838 & 0.2708 & 0.2818 & 0.5338 & 0.4399 \\
MA-GMF++ &
  0.2831$^{\dagger}$ &
  0.1453$^{\dagger}$ &
  0.3338$^{\dagger}$ &
  0.2654 &
  0.2907$^{\dagger}$ &
  0.5145 &
  0.4336$^{\dagger}$ \\ \midrule
MLP    & 0.2986 & 0.1340          & 0.4506 & 0.2869 & 0.2934 & 0.5367 & 0.4465 \\
MLP++  & 0.3170 & 0.1865          & 0.4470 & 0.3016 & 0.3100 & 0.5455 & 0.4585 \\
MA-MLP++ &
  0.3167$^{\dagger}$ &
  0.1806$^{\dagger}$ &
  0.4584 &
  0.3026 &
  0.3105$^{\dagger + *}$ &
  0.5419 &
  0.4544 \\ \midrule
NMF    & 0.3214 & 0.1717          & 0.4265 & 0.3014 & 0.2848 & 0.5430 & 0.4488 \\
NMF++ &
  0.3332 &
  0.1921 &
  \textbf{0.4595} &
  \textbf{0.3271} &
  0.3008 &
  \textbf{0.5590} &
  \textbf{0.4702} \\
MA-NMF++ &
  \textbf{0.3415$^{\dagger + *}$} &
  0.1896 &
  0.4433 &
  0.3228$^{\dagger}$ &
  \textbf{0.3158$^{\dagger \ddagger + *}$} &
  0.5573 &
  0.4578 \\ \midrule
MAML   & 0.3168 & \textbf{0.2083} & 0.4491 & 0.3152 & 0.2989 & 0.5463 & 0.4671 \\
FOREC  & 0.3040 & 0.1983          & 0.4458 & 0.3191 & 0.2927 & 0.5442 & 0.4683 \\ \bottomrule
\end{tabular}%
}
\vspace{-.2in}
\end{table}

\subsubsection{Do \ac{MA} models outperform market-unaware models when trained with the best available source?} Viewing Table \ref{tab:pairwise-experiments-best}, we first note that MA models outperform all single market variants, highlighting the utility of selecting a good source market, consistent with prior research \cite{xmrec_paper, roitero_leveraring2020}. MA models \textit{significantly} outperform single-market variants depending on the market and model, with more significant improvements seen for \acs{MA-GMF}++ (5 of 7 markets) than \acs{MA-MLP}++ (3 of 7) or \acs{MA-NMF}++ (3 of 7). Consistent improvements over the single-market models are surprisingly seen for some larger markets i.e.~ \textit{ca} and \textit{de} (but not for \textit{uk}), showing larger markets can sometimes benefit from auxiliary market data. However, the results are less consistent when comparing the MA models with their augmented but market-unaware models, especially as model size increases. MA-GMF++ improves over GMF++ in 4 of 7 markets, MA-MLP++ improves over MLP++ in 3 of 7 markets, and finally, MA-NMF++ improves over NMF++ only in 2 markets. In fact, for \textit{in}, \textit{fr}, \textit{mx} and \textit{uk}, we see that NMF++ outperforms \acs{MA-NMF}++. Furthermore, only MA-NMF++ on \textit{ca} significantly outperforms NMF++. We can thus conclude that while \ac{MA} models improve over market unaware models in some cases, \textit{selecting a source market remains an important factor for improving performance given a target market}. While this conclusion holds, we note that in general, data from multiple source markets may be unavailable, or otherwise data from target markets might be unavailable --- making best source selection unviable \cite{roitero_leveraring2020}. In such cases, results from the average-source experiments have to be considered.

\subsubsection{How do \ac{MA} models compare against meta-learning models when trained on the best source?}  We now compare MA models against MAML/ FOREC. We first note that at least one MA model beats MAML/FOREC for all markets but \textit{jp} and \textit{uk}. MA-NMF++, in particular, outperforms both MAML and FOREC for 4 of 7 markets. We see MA-NMF++ significantly outperforms both MAML/FOREC for \textit{de} and \textit{ca}. MAML achieves the best performance for \textit{jp}, beating other models by a large margin. In conclusion, we observe  similar performance of our MA models compared to meta-learning models, while outperforming them in some cases. This again indicates the effectiveness of our market embedding layer, especially when the training times are considered, which we discuss next.

\begin{figure}[t!]
    \centering

    \vspace{-.1in}
    \includegraphics[width=\textwidth]{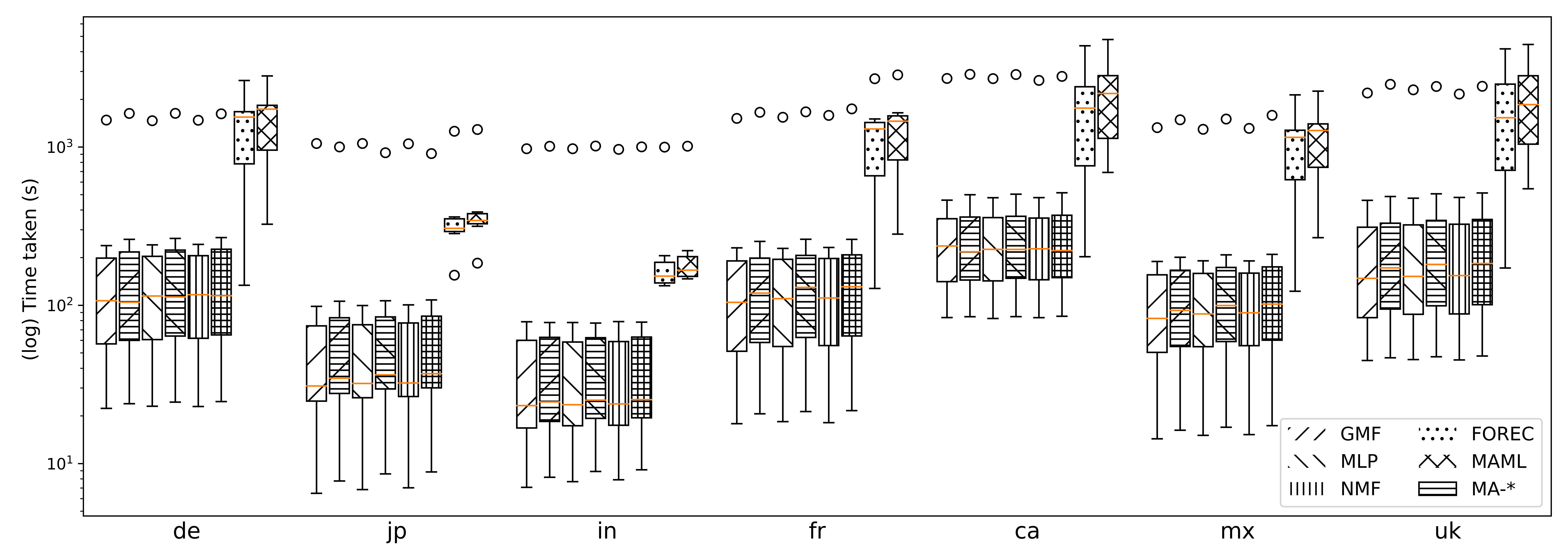}
    \caption{Time taken to train a model for a target market across all source markets, where time is on a log scale. MA and market-unaware models require similar training times, while meta-learning models require significantly more.}
    \label{fig:time_taken}
    \vspace{-.3in}
\end{figure}

\subsubsection{How do training times compare across models? Are \ac{MA} models time-efficient?} We plot the time taken to train all models for a given target market (distributed across the seven different source markets) in Figure \ref{fig:time_taken}, where the time taken is on a log scale. From this, we can see that the meta-learning models take far longer to train compared to \ac{MA} models. We note that \ac{MA} models require \textit{only} 15\% of the time taken to train meta-learning models, with \ac{MA} models requiring about the same time to train as market-unaware models. This is unsurprising, since MAML requires an inner and outer loop, as well as requiring the expensive computation of second-order derivatives \cite{antoniou2019_howto, finn2017model}. FOREC uses MAML in addition to fine-tuning the target market, so training FOREC takes up even more training time. In conclusion, \ac{MA} models achieve better or similar performance to MAML/FOREC while requiring much less training time. 

\partitle{Discussion.} Overall, we can conclude for \avgsrc that \ac{MA} models outperform both market-unaware baselines as well as meta-learning models, demonstrating the effectiveness of MA models across multiple sources. For \bestsrc i.e.~ when best-source selection is viable, the results are mixed: MA models always outperform single model variants; they outperform market-unaware models for many, but not all, markets; and an MA model either matches or outperforms meta-learning models for all markets. 

A fair question to ask is whether an increase in performance of \ac{MA} over market-unaware models can be attributed to the increase in the number of parameters from the market embeddings. However, this increase is minuscule compared to model sizes, especially for \ac{NMF} and \ac{MLP} i.e.~ for $t$ markets and $D$ dimensional user/item/market embeddings, the increase is just $tD$ parameters. In the pairwise experiments, this difference is just $16 (= 2 * 8)$, much fewer than $19929$, the number of parameters of a MLP model for the smallest target/source pair (\textit{in}/\textit{jp}).

While meta-learning models implicitly model the market during training, MA models show that this may be insufficient. We attribute the success of MA models to this explicit modelling of the markets: by adapting item representations depending on the market, the model may be better able to distinguish between recommendation in different markets more than market-unaware and meta-learning models. As we observe a better performance on \avgsrc, we can \resulthl{conclude for \textbf{RQ1}, that market-aware models exhibit a more robust performance compared to other models either matching or outperforming baselines in many settings}. While this indicates that market-aware models are more effective models in general, in some cases meta-learning models seem to learn better from the most suitable market: in these cases, MA models achieve similar performance. However, it is critical to note that MA models achieve this while requiring far less computational power. Moreover, it is evident that \ac{MA} models  do not add much complexity to non-\ac{MA} models, while empowering the model to capture the market's attributes more effectively, resulting in an efficient and effective model.

\subsection{Global Experiments}\label{sec:exp_single_model}

Table \ref{tab:global-model} reports the results of training one global recommendation model for all markets. We see that \ac{MA} models outperform baselines in many cases, even beating meta-learning models for almost all markets.

\begin{table*}[t!]
\centering
\caption{Global experiments: All markets are trained in unison. Best model for a market is in \textbf{bold}. Significance test ($p<\frac{0.05}{9}$) results are also reported comparing MA models with market unaware ($\ddagger$), MAML ($*$) and FOREC ($+$). }
\label{tab:global-model}
\resizebox{\columnwidth}{!}{%
\begin{tabular}{@{}llllllll@{}}
\toprule
\textbf{Method} &
  \multicolumn{1}{c}{\textbf{de}} &
  \multicolumn{1}{c}{\textbf{jp}} &
  \multicolumn{1}{c}{\textbf{in}} &
  \multicolumn{1}{c}{\textbf{fr}} &
  \multicolumn{1}{c}{\textbf{ca}} &
  \multicolumn{1}{c}{\textbf{mx}} &
  \multicolumn{1}{c}{\textbf{uk}} \\ \midrule
GMF++    & 0.3166 & 0.1781 & 0.4535          & 0.2884 & 0.2921                            & 0.5245 & 0.4481 \\
MA-GMF++ & 0.3073 & 0.1817 & 0.4554          & 0.2836 & 0.3015$^{ *}$                     & 0.5262 & 0.4504 \\ \midrule
MLP++    & 0.3268 & 0.2127 & 0.4479          & 0.2953 & 0.3048                            & 0.5376 & 0.4491 \\
MA-MLP++ & 0.3158 & 0.2195 & 0.4398          & 0.2958 & \textbf{0.3178$^{ \ddagger + *}$} & 0.5258 & 0.4535 \\ \midrule
NMF++    & 0.3262 & 0.1930 & \textbf{0.4796} & 0.3030 & 0.2851                            & 0.5340 & 0.4476 \\
MA-NMF++ &
  \textbf{0.3442$^{ \ddagger +}$} &
  \textbf{0.2212} &
  0.4602 &
  \textbf{0.3052} &
  0.3112$^{ \ddagger + *}$ &
  \textbf{0.5536$^{ \ddagger *}$} &
  \textbf{0.4604$^{ \ddagger + *}$} \\ \midrule
MAML     & 0.3281 & 0.1860 & 0.4736          & 0.3022 & 0.2836                            & 0.5317 & 0.4474 \\
FOREC    & 0.3249 & 0.1956 & 0.4778          & 0.3033 & 0.2947                            & 0.5409 & 0.4474 \\ \bottomrule
\end{tabular}%
}
\vspace{-.2in}
\end{table*}

\subsubsection{How do \ac{MA} models compare with market-unaware models?} MA-variant models outperform market-unaware models in 15 of 21 settings, but results differ across models: MA-GMF++ (5 of 7), MA-MLP++ (4 of 7) and MA-NMF++ (6 of 7). MA-MLP++ significantly outperforms MLP++ for \textit{ca} whereas MA-NMF++ significantly outperforms NMF++ for four markets. We also note that MA models for the largest markets, \textit{uk} and \textit{ca}, outperform both market-unaware and meta-learning models. We observe mixed results for smaller markets: for \textit{jp}, MA consistently improves over market-unaware variants, but for \textit{in}, only MA-GMF++ outperforms GMF++. Overall, we can conclude that \textit{MA} models outperform market-aware models in several settings, especially for larger markets and models. 

\subsubsection{How do \ac{MA} models compare with meta-learning-based models?}
We first note that an MA model (typically MA-NMF++) beats MAML/FOREC for all markets except \textit{in}. Indeed, MA-NMF++ beats \textit{both} MAML and FOREC for all markets except \textit{in}. It \textit{significantly} outperforms MAML for \textit{ca}, \textit{mx} and \textit{uk} markets, and FOREC for \textit{de}, \textit{ca} and \textit{uk} --- the larger markets. For \textit{ca}, we see all three MA models significantly outperform MAML, with MA-MLP++ and MA-GMF++ significantly outperforming FOREC. On the whole, we see that in a global setting, MA models outperform meta-learning methods in nearly all markets, and in particular the larger markets.

\partitle{Discussion.} We can conclude for \textbf{RQ3} that \ac{MA} models are more suitable than market unaware or meta-learning models if a global model is used for recommendation across all markets. This is critical for cases where various markets exist, empowering the model to take advantage of various user behaviours across different markets to improve recommendation in the target market. Moreover, it also leaves the problem of selecting the `best source' to the model (i.e.~ the market embedding), as the model consumes the whole data and synthesises knowledge from multiple markets. MA models seem to have an advantage over market-unaware and meta-learning models, especially for larger markets. This is likely due to the market embedding, allowing markets to distinguish source- and target-market behaviours. As more data is collected, MA models, which perform better in the global setting for larger markets, are likely to have a clear advantage.

\section{Conclusions \& Future Work}

In this work, we proposed simple yet effective \ac{MA} models for the \ac{CMR} task. In a \textit{pairwise} setting where models are trained with a single source market, \ac{MA} models on average outperform baselines in most settings, showcasing their robustness. Considering the best source market, we showed that MA models match or outperform baselines for many markets. We showed that they require far less time to train compared to meta-learning models. Next, we trained a global model for all markets and showed that \ac{MA} models match or outperform market-unaware models in nearly all settings, and outperform meta-learning models for all but one market. For future work, we plan to experiment with more complex \ac{MA} models in a limited data setting.  We also plan to investigate the utility of \ac{MA} models in a zero-shot setting, substituting the market-embedding of the new market with a similar market. In addition, we want to consider data selection techniques, since we speculate that not all data from a given source market will be useful for a given target market.

\subsubsection{Acknowledgements} 

The authors would like to thank Hamed Bonab for help with code and data, and Ruben van Heusden for help with statistical testing. The authors also thank the reviewers for their valuable feedback. This research was supported by the NWO Innovational Research Incentives Scheme Vidi (016.Vidi.189.039), the NWO Smart Culture - Big Data / Digital Humanities (314-99-301), and the H2020-EU.3.4. - SOCIETAL CHALLENGES - Smart, Green And Integrated Transport (814961). All content represents the opinion of the authors, which is not necessarily shared or endorsed by their respective employers and/or sponsors.

\renewcommand\bibname{References}
\bibliographystyle{plainnat}
\bibliography{references}

\begin{thebibliography}{27}
\providecommand{\natexlab}[1]{#1}
\providecommand{\url}[1]{\texttt{#1}}
\expandafter\ifx\csname urlstyle\endcsname\relax
  \providecommand{\doi}[1]{doi: #1}\else
  \providecommand{\doi}{doi: \begingroup \urlstyle{rm}\Url}\fi

\bibitem[Antoniou et~al.(2019)Antoniou, Edwards, and
  Storkey]{antoniou2019_howto}
Antreas Antoniou, Harrison Edwards, and Amos~J. Storkey.
\newblock How to train your {MAML}.
\newblock In \emph{7th International Conference on Learning Representations,
  {ICLR} 2019, New Orleans, LA, USA, May 6-9, 2019}. OpenReview.net, 2019.
\newblock URL \url{https://openreview.net/forum?id=HJGven05Y7}.

\bibitem[Bonab et~al.(2021)Bonab, Aliannejadi, Vardasbi, Kanoulas, and
  Allan]{xmrec_paper}
Hamed Bonab, Mohammad Aliannejadi, Ali Vardasbi, Evangelos Kanoulas, and James
  Allan.
\newblock \emph{Cross-Market Product Recommendation}, page 110–119.
\newblock Association for Computing Machinery, New York, NY, USA, 2021.
\newblock ISBN 9781450384469.
\newblock URL \url{https://doi.org/10.1145/3459637.3482493}.

\bibitem[Cao et~al.(2022)Cao, Cong, Liu, and Wang]{DBLP:conf/sigir/CaoCLW22}
Jiangxia Cao, Xin Cong, Tingwen Liu, and Bin Wang.
\newblock Item similarity mining for multi-market recommendation.
\newblock In \emph{Proceedings of the 45th International ACM SIGIR Conference
  on Research and Development in Information Retrieval}, SIGIR '22, page
  2249–2254, New York, NY, USA, 2022. Association for Computing Machinery.
\newblock ISBN 9781450387323.
\newblock \doi{10.1145/3477495.3531839}.

\bibitem[Chae et~al.(2020)Chae, Kim, Chau, and Kim]{chae_arcf2020}
Dong-Kyu Chae, Jihoo Kim, Duen~Horng Chau, and Sang-Wook Kim.
\newblock Ar-cf: Augmenting virtual users and items in collaborative filtering
  for addressing cold-start problems.
\newblock In \emph{Proceedings of the 43rd International ACM SIGIR Conference
  on Research and Development in Information Retrieval}, SIGIR '20, page
  1251–1260, New York, NY, USA, 2020. Association for Computing Machinery.
\newblock ISBN 9781450380164.
\newblock \doi{10.1145/3397271.3401038}.

\bibitem[Cheng et~al.(2016)Cheng, Koc, Harmsen, Shaked, Chandra, Aradhye,
  Anderson, Corrado, Chai, Ispir, Anil, Haque, Hong, Jain, Liu, and
  Shah]{cheng2016wide}
Heng-Tze Cheng, Levent Koc, Jeremiah Harmsen, Tal Shaked, Tushar Chandra,
  Hrishi Aradhye, Glen Anderson, Greg Corrado, Wei Chai, Mustafa Ispir, Rohan
  Anil, Zakaria Haque, Lichan Hong, Vihan Jain, Xiaobing Liu, and Hemal Shah.
\newblock Wide \& deep learning for recommender systems.
\newblock In \emph{Proceedings of the 1st Workshop on Deep Learning for
  Recommender Systems}, DLRS 2016, page 7–10, New York, NY, USA, 2016.
  Association for Computing Machinery.
\newblock ISBN 9781450347952.
\newblock \doi{10.1145/2988450.2988454}.

\bibitem[Elkahky et~al.(2015)Elkahky, Song, and He]{elkahky_multivew2015}
Ali~Mamdouh Elkahky, Yang Song, and Xiaodong He.
\newblock A multi-view deep learning approach for cross domain user modeling in
  recommendation systems.
\newblock In \emph{Proceedings of the 24th International Conference on World
  Wide Web}, WWW '15, page 278–288, Republic and Canton of Geneva, CHE, 2015.
  International World Wide Web Conferences Steering Committee.
\newblock ISBN 9781450334693.
\newblock \doi{10.1145/2736277.2741667}.

\bibitem[Ferwerda et~al.(2016)Ferwerda, Vall, Tkalcic, and
  Schedl]{ferwerda2016exploring}
Bruce Ferwerda, Andreu Vall, Marko Tkalcic, and Markus Schedl.
\newblock Exploring music diversity needs across countries.
\newblock In \emph{Proceedings of the 2016 Conference on User Modeling
  Adaptation and Personalization}, UMAP '16, page 287–288, New York, NY, USA,
  2016. Association for Computing Machinery.
\newblock ISBN 9781450343688.
\newblock \doi{10.1145/2930238.2930262}.

\bibitem[Finn et~al.(2017)Finn, Abbeel, and Levine]{finn2017model}
Chelsea Finn, Pieter Abbeel, and Sergey Levine.
\newblock Model-agnostic meta-learning for fast adaptation of deep networks.
\newblock In \emph{Proceedings of the 34th International Conference on Machine
  Learning - Volume 70}, ICML'17, page 1126–1135. JMLR.org, 2017.

\bibitem[Ganin and Lempitsky(2015)]{ganin2015unsupervised}
Yaroslav Ganin and Victor Lempitsky.
\newblock Unsupervised domain adaptation by backpropagation.
\newblock In Francis Bach and David Blei, editors, \emph{Proceedings of the
  32nd International Conference on Machine Learning}, volume~37 of
  \emph{Proceedings of Machine Learning Research}, pages 1180--1189, Lille,
  France, 07--09 Jul 2015. PMLR.
\newblock URL \url{https://proceedings.mlr.press/v37/ganin15.html}.

\bibitem[Ge et~al.(2020)Ge, Xu, Liu, Fu, Sun, and Zhang]{ge2020learning}
Yingqiang Ge, Shuyuan Xu, Shuchang Liu, Zuohui Fu, Fei Sun, and Yongfeng Zhang.
\newblock Learning personalized risk preferences for recommendation.
\newblock In \emph{Proceedings of the 43rd International ACM SIGIR Conference
  on Research and Development in Information Retrieval}, SIGIR '20, page
  409–418, New York, NY, USA, 2020. Association for Computing Machinery.
\newblock ISBN 9781450380164.
\newblock \doi{10.1145/3397271.3401056}.

\bibitem[He et~al.(2017)He, Liao, Zhang, Nie, Hu, and Chua]{he2017ncf}
Xiangnan He, Lizi Liao, Hanwang Zhang, Liqiang Nie, Xia Hu, and Tat-Seng Chua.
\newblock Neural collaborative filtering.
\newblock In \emph{Proceedings of the 26th International Conference on World
  Wide Web}, WWW '17, page 173–182, Republic and Canton of Geneva, CHE, 2017.
  International World Wide Web Conferences Steering Committee.
\newblock ISBN 9781450349130.
\newblock \doi{10.1145/3038912.3052569}.

\bibitem[Hu et~al.(2018)Hu, Zhang, and Yang]{hu2018conet}
Guangneng Hu, Yu~Zhang, and Qiang Yang.
\newblock Conet: Collaborative cross networks for cross-domain recommendation.
\newblock In \emph{Proceedings of the 27th ACM International Conference on
  Information and Knowledge Management}, CIKM '18, page 667–676, New York,
  NY, USA, 2018. Association for Computing Machinery.
\newblock ISBN 9781450360142.
\newblock \doi{10.1145/3269206.3271684}.

\bibitem[Huang et~al.(2013)Huang, He, Gao, Deng, Acero, and
  Heck]{huang_dssm2013}
Po-Sen Huang, Xiaodong He, Jianfeng Gao, Li~Deng, Alex Acero, and Larry Heck.
\newblock Learning deep structured semantic models for web search using
  clickthrough data.
\newblock In \emph{Proceedings of the 22nd ACM International Conference on
  Information \& Knowledge Management}, CIKM '13, page 2333–2338, New York,
  NY, USA, 2013. Association for Computing Machinery.
\newblock ISBN 9781450322638.
\newblock \doi{10.1145/2505515.2505665}.

\bibitem[Im and Hars(2007)]{im2007does}
Il~Im and Alexander Hars.
\newblock Does a one-size recommendation system fit all? the effectiveness of
  collaborative filtering based recommendation systems across different domains
  and search modes.
\newblock \emph{ACM Trans. Inf. Syst.}, 26\penalty0 (1):\penalty0 4–es, nov
  2007.
\newblock ISSN 1046-8188.
\newblock \doi{10.1145/1292591.1292595}.

\bibitem[Kanagawa et~al.(2019)Kanagawa, Kobayashi, Shimizu, Tagami, and
  Suzuki]{kanagawa2019cross}
Heishiro Kanagawa, Hayato Kobayashi, Nobuyuki Shimizu, Yukihiro Tagami, and
  Taiji Suzuki.
\newblock Cross-domain recommendation via deep domain adaptation.
\newblock In Leif Azzopardi, Benno Stein, Norbert Fuhr, Philipp Mayr, Claudia
  Hauff, and Djoerd Hiemstra, editors, \emph{Advances in Information
  Retrieval}, pages 20--29, Cham, 2019. Springer International Publishing.
\newblock ISBN 978-3-030-15719-7.
\newblock \doi{10.1007/978-3-030-15719-7_3}.

\bibitem[Kang and McAuley(2018)]{kang2018self}
Wang-Cheng Kang and Julian McAuley.
\newblock Self-attentive sequential recommendation.
\newblock In \emph{ICDM}, pages 197--206, 2018.

\bibitem[Krishnan et~al.(2020)Krishnan, Das, Bendre, Yang, and
  Sundaram]{krishnan_ctxlinvariants20202}
Adit Krishnan, Mahashweta Das, Mangesh Bendre, Hao Yang, and Hari Sundaram.
\newblock Transfer learning via contextual invariants for one-to-many
  cross-domain recommendation.
\newblock In \emph{Proceedings of the 43rd International ACM SIGIR Conference
  on Research and Development in Information Retrieval}, SIGIR '20, page
  1081–1090, New York, NY, USA, 2020. Association for Computing Machinery.
\newblock ISBN 9781450380164.
\newblock \doi{10.1145/3397271.3401078}.

\bibitem[Li and Tuzhilin(2020)]{li2020ddtcdr}
Pan Li and Alexander Tuzhilin.
\newblock Ddtcdr: Deep dual transfer cross domain recommendation.
\newblock In \emph{Proceedings of the 13th International Conference on Web
  Search and Data Mining}, WSDM '20, page 331–339, New York, NY, USA, 2020.
  Association for Computing Machinery.
\newblock ISBN 9781450368223.
\newblock \doi{10.1145/3336191.3371793}.

\bibitem[Li et~al.(2020)Li, Xu, Zhao, Fang, Chen, and Zhao]{li2020atlrec}
Ying Li, Jia-Jie Xu, Peng-Peng Zhao, Jun-Hua Fang, Wei Chen, and Lei Zhao.
\newblock Atlrec: An attentional adversarial transfer learning network for
  cross-domain recommendation.
\newblock \emph{Journal of Computer Science and Technology}, 35\penalty0
  (4):\penalty0 794--808, 2020.
\newblock \doi{10.1007/s11390-020-0314-8}.

\bibitem[Lu et~al.()Lu, Zhong, Zhao, Xiang, Pan, and Yang]{lu2013selective}
Zhongqi Lu, Erheng Zhong, Lili Zhao, Evan~Wei Xiang, Weike Pan, and Qiang Yang.
\newblock \emph{Selective Transfer Learning for Cross Domain Recommendation},
  pages 641--649.
\newblock \doi{10.1137/1.9781611972832.71}.

\bibitem[Mirbakhsh and Ling(2015)]{mirbakhsh_improving2015}
Nima Mirbakhsh and Charles~X. Ling.
\newblock Improving top-n recommendation for cold-start users via cross-domain
  information.
\newblock \emph{ACM Trans. Knowl. Discov. Data}, 9\penalty0 (4), jun 2015.
\newblock ISSN 1556-4681.
\newblock \doi{10.1145/2724720}.

\bibitem[Perera and Zimmermann(2019)]{perera2019cngan}
Dilruk Perera and Roger Zimmermann.
\newblock Cngan: Generative adversarial networks for cross-network user
  preference generation for non-overlapped users.
\newblock In \emph{The World Wide Web Conference}, WWW '19, page 3144–3150,
  New York, NY, USA, 2019. Association for Computing Machinery.
\newblock ISBN 9781450366748.
\newblock \doi{10.1145/3308558.3313733}.

\bibitem[Rafailidis and Crestani(2017)]{rafailidis_collab2017}
Dimitrios Rafailidis and Fabio Crestani.
\newblock A collaborative ranking model for cross-domain recommendations.
\newblock In \emph{Proceedings of the 2017 ACM on Conference on Information and
  Knowledge Management}, CIKM '17, page 2263–2266, New York, NY, USA, 2017.
  Association for Computing Machinery.
\newblock ISBN 9781450349185.
\newblock \doi{10.1145/3132847.3133107}.

\bibitem[Roitero et~al.(2020)Roitero, Carterrete, Mehrotra, and
  Lalmas]{roitero_leveraring2020}
Kevin Roitero, Ben Carterrete, Rishabh Mehrotra, and Mounia Lalmas.
\newblock Leveraging behavioral heterogeneity across markets for cross-market
  training of recommender systems.
\newblock In \emph{Companion Proceedings of the Web Conference 2020}, WWW '20,
  page 694–702, New York, NY, USA, 2020. Association for Computing Machinery.
\newblock ISBN 9781450370240.
\newblock \doi{10.1145/3366424.3384362}.

\bibitem[Wang et~al.(2020)Wang, Niepert, and Li]{wang2019recsys}
Cheng Wang, Mathias Niepert, and Hui Li.
\newblock Recsys-dan: Discriminative adversarial networks for cross-domain
  recommender systems.
\newblock \emph{IEEE Transactions on Neural Networks and Learning Systems},
  31\penalty0 (8):\penalty0 2731--2740, 2020.
\newblock \doi{10.1109/TNNLS.2019.2907430}.

\bibitem[Yuan et~al.(2019)Yuan, Yao, and Benatallah]{yuan2019darec}
Feng Yuan, Lina Yao, and Boualem Benatallah.
\newblock Darec: Deep domain adaptation for cross-domain recommendation via
  transferring rating patterns.
\newblock In \emph{Proceedings of the 28th International Joint Conference on
  Artificial Intelligence}, IJCAI'19, page 4227–4233. AAAI Press, 2019.
\newblock ISBN 9780999241141.

\bibitem[Zhao et~al.(2020)Zhao, Li, Xiao, Deng, and Sun]{zhao2020catn}
Cheng Zhao, Chenliang Li, Rong Xiao, Hongbo Deng, and Aixin Sun.
\newblock Catn: Cross-domain recommendation for cold-start users via aspect
  transfer network.
\newblock In \emph{Proceedings of the 43rd International ACM SIGIR Conference
  on Research and Development in Information Retrieval}, SIGIR '20, page
  229–238, New York, NY, USA, 2020. Association for Computing Machinery.
\newblock ISBN 9781450380164.
\newblock \doi{10.1145/3397271.3401169}.
\newblock URL \url{https://doi.org/10.1145/3397271.3401169}.

\end{thebibliography}

\end{document}